\documentclass[12pt]{article}
\usepackage{graphicx}
\newcommand{\bb}{\begin{equation}}
\newcommand{\ee}{\end{equation}}
\newcommand{\ba}{\begin{eqnarray}}
\newcommand{\ea}{\end{eqnarray}}

\begin{document}

\title{{\bf Can Two Ultrarelativistic Objects\\ 
Lose Almost All Their Energy\\ 
to Gravitational Radiation?}}

\author{
Don N. Page
\thanks{Internet address:
profdonpage@gmail.com}
\\
Department of Physics\\
4-183 CCIS\\
University of Alberta\\
Edmonton, Alberta T6G 2E1\\
Canada
}

\date{2022 December 7}

\maketitle
\large

\begin{abstract}
\baselineskip 18 pt

In 2007 Pretorius and Khurana \cite{Pretorius:2007jn} did ``speculate that at threshold [at a critical impact parameter], all of the kinetic energy of the system [two ultrarelativistic black holes] is converted to gravitational waves, which can be an arbitrarily large fraction of the total energy.''  However, in 2012 Sperhake, Berti, Cardoso, and Pretorius \cite{Sperhake:2012me} performed numerical calculations that led them to the contrary conclusion:  ``An extrapolation of our results to the limit $\gamma \rightarrow \infty$ suggests that about half of the center-of-mass energy of the system can be emitted in gravitational radiation, while the rest must be converted into rest-mass and spin energy.''  Here I present arguments against this latter conclusion and in support of the earlier speculation that for sufficiently large $\gamma$, all but an arbitrarily small fraction of the total energy can be radiated away.

\end{abstract}

\normalsize

\baselineskip 20.6 pt

\newpage

\section{Introduction}

LIGO has had enormous success in detecting the gravitational waves from astrophysical inspiraling black holes that coalesce to form a larger black hole \cite{LIGOScientific:2016aoc,LIGOScientific:2018mvr,LIGOScientific:2020ibl,LIGOScientific:2021djp}.  
In these cases the black holes have been inspiraling long before their gravitational wave emission is strong enough to be detected by LIGO, so that well before the final coalescence, the two black holes are moving with nonrelativistic velocities $v \ll c$ relative to each other and are gravitationally bound to each other.

Astrophysically, it seems that it would be very rare (and not yet observed) for two black holes or other macroscopic objects to approach at relativistic velocities $v \sim c$ before coalescing.  However, it is an interesting question what would happen if this occurred.  Indeed, there has been an enormous amount of work [5--92] calculating what happens for compact objects approaching each other at arbitrary relative velocities in 4-dimensional spacetime (not including considerable additional work that has also been done in higher dimensions that I shall not discuss here).  Much of this work has shown essentially that if the Schwarzschild radius corresponding to the total center-of-momentum (COM) energy $E$, that is, $R_S = 2GE/c^4 = 2E$ (where here and henceforth I shall use units with $G = c = 1$), is much larger than the sum $R$ of the intrinsic linear sizes of the two objects, then the composition of each object does not matter significantly for how much gravitational wave energy $\Delta E$ is emitted, which, for fixed total COM energy $E$ instead depends almost entirely on the impact parameter $b \equiv J/p \equiv \beta E$, where $J$ is the total angular momentum in the COM and $p$ is the magnitude of the initial spatial momentum of each object in the COM (equal magnitudes but opposite directions, since the total spatial momentum is zero in the COM).  Here
\bb
\beta \equiv \frac{b}{E} \equiv \frac{J}{pE} \approx \frac{J}{2E^2} = \frac{2J}{R_S}
\label{beta}
\ee
is a dimensionless measure of the impact parameter (not to be confused with $v/c$).
The approximation of the last two terms occurs in the limit that the COM energy $E$ is much greater than the sum $M = m_1 + m_2$ of the rest masses $m_1$ and $m_2$ of the two objects, so that then $p \approx E/2$, giving $b \approx 2J/E$.  (Note that I am {\it not} using $M$ for the total ADM mass of the spacetime as many others do, but rather $E$.)

For fixed $E \gg M$ and $R_S = 2E \gg R$, there is at least one critical impact parameter $b_c = \beta_c E = \beta_c R_S/2$ (with $\beta_c$ some dimensionless number of the order of unity, which Shibata, Okawa, and Yamamoto \cite{Shibata:2008rq} have estimated to be about $2.5/v$), that should be independent of $E$, say with value $\approx 2.5$, in the limit $E/M \rightarrow \infty$ and $R_S/R \rightarrow \infty$, such that for impact parameter $b$ infinitesimally on one side of $b_c$, the two objects will, at least eventually, form a single black hole, but for $b$ infinitesimally on the other side of $b_c$, the two objects will scatter separately to infinity (though leaving open the possibility that either of the objects might separately become a black hole even if it were not one originally, or that both might become black holes that never merge).

It is perhaps simplest to assume that for fixed $E \gg M$ and $R_S = 2E \gg R$, there is a unique critical impact parameter $b_c$ such that for all smaller impact parameters, $b < b_c$, the two objects eventually merge and form a single black hole, but for all larger impact parameters, $b > b_c$, the two objects scatter separately to infinity.  However, near the end of this paper I shall discuss the possibility that this might not be true, and that there might instead be a larger odd number of critical impact parameters $b_c$.  Counting with positive integers $n$, with $n=1$ for the smallest critical impact parameter $b_c = b_{c,1}$ (so that at all lower values of $b$, that is for all $0 \leq b < b_{c,1}$, a single black hole forms), each odd critical impact parameter, $b_{c,2n-1}$, has, for $b$ infinitesimally smaller than $b_{c,2n-1}$, the two objects merging to form a single black hole and has, for $b$ infinitesimally larger than $b_{c,2n-1}$, the two objects scattering separately to infinity.  Conversely, each even critical impact parameter, $b_{c,2n}$, has, for $b$ infinitesimally smaller than $b_{c,2n}$, the two objects scattering separately to infinity and has, for $b$ infinitesimally larger than $b_{c,2n}$, the two objects merging to form a single black hole.

In the following, when I discuss the impact parameter $b_c$, if in fact there are more than one, I shall mean by $b_c$ the particular critical impact parameter at which the largest fraction $f$ of the fixed total energy $E$ is emitted into gravitational waves of energy $\Delta E = f E$.  Similarly, $\beta_c \equiv b_c/E$ will denote the dimensionless quantity analogous to that particular critical impact parameter.

When $b$ is much smaller than $b_c = \beta_c E$, it appears that the fraction $f$ of the initial COM energy $E$ that is radiated into gravitational waves of energy $\Delta E = fE$ is significantly less than unity (approximately $0.14\pm 0.03$ for a head-on collision, $b=0$, when $E \gg M$ \cite{Sperhake:2008ga}), and it seems that increasing $b$ increases $f = \Delta E/E$.  On the other hand, when $b$ is much larger than $b_c = \beta_c E$, the two objects have a small scattering angle and also do not emit a large fraction of their energy into gravitational radiation, and $\Delta E$, the amount radiated, goes down with increasing $b$ as the scattering angle also decreases.  Therefore, there should be some value in between for $b$, say $b_m = \beta_m E$, at which the maximum fraction $f = \Delta E/E$, say $f_m$, of the total COM energy $E$ (which is to be kept fixed when $b$ is varied) is emitted as gravitational wave energy $\Delta E$.  If there are in fact more than one critical impact parameters, it also seems possible that $f$ does not decrease monotonically from $f_m$ as $b$ is moved in either direction away from $b_m$, so that there may be more than one local maximum for $f$ (perhaps one for each odd critical impact parameter $b_{c,2n-1}$), but I am taking $f_m$ to be the global maximum for $f$.  It seems plausible that $b_m$ would be close to the critical impact parameter $b_c$ (so $\beta_m$ is close to $\beta_c$), especially in the limit $E/M \rightarrow \infty$ and $R_S/R \rightarrow \infty$, but here I shall not assume that this needs to be true.

Now Conjecture 1, a slight generalization of the 2007 speculation of Pretorius and Khurana \cite{Pretorius:2007jn} to an arbitrary pair of initial objects of finite size, is that in the limit $\gamma \equiv E/M \rightarrow \infty$ and $R_S/R \rightarrow \infty$, $f_m \rightarrow 1$.

Slightly stronger is Conjecture 2, that at the critical impact parameter $b_c$ that has the largest $f = \Delta E/E$ for finite $E/M$ and finite $R_S/R$, when one takes the limit $\gamma \equiv E/M \rightarrow \infty$ and $R_S/R \rightarrow \infty$, $f = \Delta E/E$ at this $\gamma$-dependent $b_c$ approaches 1.

However, in 2012 Sperhake, Berti, Cardoso, and Pretorius (SBCP) \cite{Sperhake:2012me} performed numerical calculations that led them to the contrary conclusion:  ``An extrapolation of our results to the limit $\gamma \rightarrow \infty$ suggests that about half of the center-of-mass energy of the system can be emitted in gravitational radiation, while the rest must be converted into rest-mass and spin energy.''

Here I wish to revisit the question of what fraction $f = \Delta E/E$ of the initial COM energy $E$ can be emitted in gravitational wave energy $\Delta E$ in the limit that $\gamma \equiv E/M$ is taken to infinity.  Despite the numerical evidence presented by \cite{Sperhake:2012me}, I shall argue that it is more plausible that $f$ can be made arbitrarily near 1 by taking $\gamma$ to infinity while fine tuning the impact parameter $b$ appropriately.

\section{Ultrarelativistic Collisions of Black Holes}

My arguments should generalize to apply for any pair of objects with total linear size $R$ and with $M$ the sum of their rest masses (here {\it not} the total ADM mass of the spacetime, which is what I shall denote by the total energy $E$), when one separately boosts the two objects so that the center-of-momentum (COM) energy $E$ is much greater than both $R$ and $M$ (in units with $G = c = 1$).  However, for simplicity and concreteness let me assume that the two objects are initially Schwarzschild (nonrotating) black holes having positive rest masses $m_1$ and $m_2$, with $M = m_1 + m_2$ and $R = 2m_1 + 2m_2 = 2M$ if one uses the Schwarzschild radius of each black hole as its linear size.

Then the critical impact parameter that gives the largest $f = \Delta E/E$ out of all critical impact parameters (if there is more than one) should be 
\bb
b_c = b_c(m_1,m_2,E) = \beta_2(m_1/E,m_2/E)E,
\label{beta2}
\ee
with the dimensionless function $\beta_2(m_1/E,m_2/E)$, a symmetric function of its two dimensionless arguments, going to a constant, which I shall call $\beta_0$, in the limit that $m_1/E \rightarrow 0$ and $m_2/E \rightarrow 0$ (which is implied by the single limit $M/E \rightarrow 0$):
\bb
\beta_0 \equiv \lim_{M/E \rightarrow 0}\frac{b_c}{E}.
\label{beta_0}
\ee 

For impact parameter $b$ just infinitesimally larger than this particular critical impact parameter $b_c$, the two black holes (each initially much smaller than the impact parameter, since their Schwarzschild radii, $2m_1$ and $2m_2$, are much smaller than $b_c \sim E$, leaving out the dimensionless factor $\beta_2(m_1/E,m_2/E)$ that is expected to be of the order of unity) do not merge into a larger black hole.  On the other hand, for impact parameter $b$ just infinitesimally smaller than the critical impact parameter $b_c$, the two black holes would merge into a larger black hole.

In this example, there are four independent quantities that with $G = c = 1$ have the dimension of mass or length or time, namely $m_1$, $m_2$, $E$, and $b$, and hence there are three independent dimensionless ratios, namely $m_1/E$, $m_2/E$, and $b/E$.  At the critical impact parameter $b_c$ given by Eq.\ (\ref{beta2}), there are only two independent dimensionless ratios, namely $m_1/E$ and $m_2/E$.

It is also interesting to restrict to an even simpler model in which each initially Schwarzschild black hole has the same mass, $m_1 = m_2 = M/2$, so that then there are only three independent quantities that have the dimension of mass or length or time, namely $M$, $E$, and $b$, and hence there are only two independent dimensionless ratios, namely $\gamma\equiv E/M$ and $\beta \equiv b/E$ (here {\it not} equal to $v/c = \sqrt{1-1/\gamma^2}$).  At the particular critical impact parameter $b_c$ that gives the maximum $f = \Delta E/E$ out of all critical impact parameters if there are more than one, which now has the form
\bb
b_c = b_c(E,M/2,M/2) = \beta_1(\gamma)E,
\label{beta1}
\ee
there is only one independent dimensionless ratio, namely $\gamma\equiv E/M$.  The fraction $f = \Delta E/E$ of the initial energy $E$ that is radiated into gravitational waves with energy $\Delta E$ at the critical impact parameter would then be a function only of $\gamma$, and the question under dispute is how $f(\gamma)$ depends on $\gamma$ in the limit that $\gamma$ is taken to infinity.

\baselineskip 21 pt

\section{Qualitative Argument for Ultrarelativistic Scattering Near the Critical Impact Parameter}

Here I shall restrict to the simplest model, with two initially incoming Schwarzschild black holes, each of mass $M/2$, which in the COM have opposite spatial momenta each of magnitude $p = (M/2)\gamma v$ with equal gamma-factors, $\gamma = 1/\sqrt{1-v^2}$, so that the total energy in the COM frame is $E = M\gamma$, and with total angular momentum $J = b p$ with impact parameter $b = J/p = 2J/(E v)$ that is near the critical impact parameter $b_c = \beta_1(\gamma)E$ with the maximum $f = \Delta E/E$ (out of all critical impact parameters, if there are more than one, not out of all possible values of the impact parameter $b$).  I shall describe the picture Pretorius and Khurana \cite{Pretorius:2007jn} painted and then argue that it seems generally qualitatively correct, despite the opposing arguments of Sperhake, Berti, Cardoso, and Pretorius (SBCP) \cite{Sperhake:2012me}.

In particular, the idea is that near the critical impact parameter $b_c = \beta_1(\gamma)E$, the two black holes, each of size much smaller than the critical impact parameter $b_c$ when $\gamma = E/M \gg 1$ as I shall always assume, spiral inward while radiating away almost all their initial energy, until they coalesce at the center to form a black hole that on a logarithmic scale has a mass closer to the original total rest mass $M$ than to the initial COM energy $E = M\gamma$.  At each time $t$ that is measured by a static observer in the COM frame at radial infinity, let $r(t)$ be some measure of the distance of each black hole from the center of momentum, such as the proper distance along a spacelike geodesic from the center of momentum that at this center is orthogonal to the worldline of the center of momentum.

With the black holes orbiting around the COM at very nearly the speed of light, one would expect, in analogy with the quadrupole formula, that the sum of the rest masses and kinetic energies of the black holes, which I shall call $E_h$ (not including the gravitational potential energy, which would be expected to be much smaller than $E_h$ for the ultrarelativistic motion, and not including the gravitational radiation energy $E_{GW} \approx E - E_h$, which would carry away significant amounts of energy to cause $E_h$ to decrease) would decrease at a rate per time that would be roughly proportional to $(E_h/r)^2$, assuming that the shape of the orbit is nearly a self-similar equiangular spiral.  Dividing this radiation power outflow from the two black holes by the negative radial velocity $v_r = dr/dt$, which would be expected to have a magnitude 1--2 orders of magnitude smaller than unity but still be of the general order of unity (i.e., not having any strong dependence on $\gamma$ when $\gamma \gg 1$), and including an unknown numerical factor of $\alpha$, one gets (here initially ignoring the effect of the absorption of gravitational waves by the black holes)
\bb
\frac{dE_h}{dr} \sim \frac{1}{\alpha}\frac{E_h^2}{r^2},
\label{dE/dr1}
\ee
which leads to a solution that asymptotically, as $r$ decreases far below the original impact parameter $b$, has the form $E_h \sim \alpha r$ for the total rest mass plus kinetic energy of the two black holes.

Sperhake, Berti, Cardoso, and Pretorius \cite{Sperhake:2012me} argue that, even in the limit $\gamma \rightarrow \infty$, a significant fraction of the kinetic energy of the two black holes is not radiated away but falls into the black holes to increase their sizes significantly.  Here I shall argue that the fraction of the total energy that falls into the black holes goes to zero in the limit that one takes $\gamma$ to infinity, so long as at each $\gamma$ one chooses the impact parameter $b$ to be near the critical impact parameter $b_c$ giving the largest $f = \Delta E/E$ out of all critical impact parameters (if there are more than one).  Alternatively, one could choose $b$ to be near $b_m$, the value of $b$ that maximizes $f$ without restricting to critical impact parameters, though I would expect $b_m$ to be near $b_c$ for each value of $\gamma$, at least for $\gamma \gg 1$.   

The argument is that during most of the inspiral, when most of the energy is radiated into gravitational waves, the black holes, initially each of mass $m = M/2 = E/(2\gamma) \ll b_c \sim E$ and hence each of Schwarzschild radii $2m = M = E/\gamma$, are so much smaller than the region of linear size $\sim r \sim E_h$ where the gravitational wave energy is localized near the black hole orbits before it flows outward, that only a tiny fraction of this energy will be absorbed by the black holes.

Let me make a crude model to estimate the order of magnitude of the fraction $F \equiv 1-f$ of the initial total COM energy $E$ that is absorbed by the black holes, as a function of $\gamma = E/M$ when it is large, and when $b$ is sufficiently close to either $b_c$ or to $b_m$ for that $\gamma$.  Each black hole of rest mass $m$ has an effective cross section for absorbing gravitational radiation that is $m^2$ multiplied by some dimensionless number (depending on the frequency spectrum of the gravitational radiation in the frame of the black hole and which is $27\pi$ for gravitational radiation of wavelength very short compared with the Schwarzschild radius $2m$ of the black hole when it is nonrotating, which is the geometric optics limit in which null geodesics fall into a Schwarzschild black hole when they have impact parameter $b < \sqrt{27}m$ \cite{Hagihara,Darwin}).  Therefore, the rate of gravitational wave energy absorbed by each black hole is some dimensionless number multiplying $m^2\rho$, where $\rho$ is the energy density of the gravitational waves at the location of the black hole.  One can crudely approximate $\rho$ to be the energy in gravitational waves within the distance $r$ of the COM origin that is the distance each black hole is from that origin at the time $t$ of absorption, divided by the volume $\sim 4\pi r^3/3$ inside that region.  Since the gravitational wave energy is produced at a rate $\approx -dE_h/dt \sim dE_h/dr$ and is flowing generally outward with an effective radial velocity near the speed of light ($c=1$), the energy within the distance $r$ of the COM origin is of the order of $r dE_h/dr$, giving an effective gravitational wave energy density $\rho \sim (r dE_h/dr)/r^3 = (1/r^2)dE_h/dr$, dropping factors of the order of unity such as $4\pi/3$ and various velocities in units of the speed of light.

\newpage

For the total rate of the increase of $E_h$, the time-dependent the rest mass plus kinetic energy of the two black holes, if we now add the positive contribution of the absorption of gravitational waves to the negative contribution from the emission of the gravitational waves, and divide by the negative $v_r = dr/dt$, we get, as an improvement over Eq.\ (\ref{dE/dr1}),
\bb
\frac{dE_h}{dr} \sim \frac{1}{\alpha}\frac{E_h^2}{r^2} - \delta\frac{m^2}{r^2}\frac{dE_h}{dr},
\label{dE/dr2}
\ee
where I have introduced a second unknown dimensionless factor $\delta$ to take into account the unknown factors of the order of unity that were dropped in the previous paragraph.  Of course, Eq.\ (\ref{dE/dr2}) can readily be rearranged to give
\bb
\frac{dE_h}{dr} \sim \left(1+\delta\frac{m^2}{r^2}\right)^{-1}\frac{1}{\alpha}\frac{E_h^2}{r^2}.
\label{dE/dr3}
\ee

Next, we need to get an approximate differential equation for how $m$ evolves as $r$ decreases.  If each black hole were absorbing radiation whose center-of-momentum energy were at rest in the COM frame (i.e., if there were no energy flux in that frame), the mass would increase at a rate that is $\gamma_h \equiv (1/2)E_h/m$ times the rate of its energy increase from the absorption of gravitational waves that was approximated above by $-(m^2/r^2)(dE_h/dt)$, so then
\bb
\frac{dm}{dt} \sim -\frac{1}{2}\frac{E_h}{m}\frac{m^2}{r^2}\frac{dE_h}{dt}.
\label{dm/dt}
\ee
Replacing the factor of $1/2$ by a third unknown dimensionless factor of $\epsilon$ to take into account the uncertainty in the mean effective velocity of the gravitational waves, rearranging Eq.\ (\ref{dm/dt}), and combining it with Eq.\ (\ref{dE/dr3}) gives
\bb
\frac{r}{m}\frac{dm}{dr} \sim -\epsilon \frac{E_h}{r}\frac{dE_h}{dr} \sim -\left(1+\delta\frac{m^2}{r^2}\right)^{-1} \frac{\epsilon}{\alpha} \frac{E_h^3}{r^3}.
\label{dlnm/dlnr}
\ee
One can then combine Eqs.\ (\ref{dE/dr3}) and (\ref{dlnm/dlnr}) to get the rate at which the logarithm of $m$ increases when the logarithm of $E_h$ decreases:
\bb
\frac{d\ln{m}}{d\ln{E_h}} \equiv \frac{E_h}{m}\frac{dm}{dE_h} \sim - \epsilon \frac{E_h^2}{r^2}.
\label{dlnm/dlnE}
\ee

One can see from Eq.\ (\ref{dE/dr3}) that when $\delta m^2/r^2 \ll 1$, as $r$ decreases with time as the black holes spiral closer together, say with initial conditions $E_h \sim E$ at $r \sim r_0$,
\bb
E_h \sim \frac{\alpha r}{1 + \alpha r/E - r/r_0},
\label{E_h}
\ee
so that as $r$ drops far below $r_0$ and $E$, $E_h$ approaches $\alpha r$.  Therefore, during this stage of the evolution, when $E_h \sim \alpha r$, Eq.\ (\ref{dlnm/dlnE}) gives
\bb
\frac{d\ln{m}}{d\ln{E_h}} \sim - \epsilon \alpha^2,
\label{dlnm/dlnE2}
\ee
so
\bb
\frac{m}{m_0} \sim \left(\frac{E_h}{E}\right)^{-\alpha^2\epsilon} 
\sim \left(\frac{\alpha r}{E}\right)^{-\alpha^2\epsilon},
\label{m/m0}
\ee
where $m_0 = M/2 = E/(2\gamma)$ is the initial rest mass of each black hole (with the sum of the two initial black hole rest masses being $2m_0 = M = E/\gamma$), and where the initial value of $E_h$, the 
total rest mass plus kinetic energies of the two black holes, is the constant total COM energy $E = \gamma M = 2\gamma m_0$, and where the initial value of $r$ at the beginning of inspiral is $r_0 \sim b \sim E$, assuming $b \sim b_c = \beta E$ with $\beta \sim \beta_c \sim 1$.

The black holes will merge when $m \sim r$, so that
\bb
1 \sim \frac{m}{r} = m_0\frac{1}{r}\frac{m}{m_0} 
\sim \frac{E}{2\gamma}\frac{\alpha}{E_h}\left(\frac{E_h}{E}\right)^{-\alpha^2\epsilon} 
= \frac{\alpha}{2\gamma}\left(\frac{E_h}{E}\right)^{-1-\alpha^2\epsilon}.
\label{m/r}
\ee
Therefore, one gets that the fraction $F$ of the initial total COM energy $R$ that is {\it not} radiated into gravitational waves is approximately
\bb
F \equiv 1-f \sim \frac{E_h}{E} \sim \left(\frac{\alpha}{2\gamma}\right)^{\frac{1}{1+\alpha^2\epsilon}} \sim \left(\frac{\alpha}{2\gamma}\right)^p,
\label{1-f}
\ee
with exponent
\bb
p \sim \frac{1}{1+\alpha^2\epsilon}.
\label{p}
\ee
Therefore, for large $\gamma \equiv E/M = E/(2m_0)$, $F = 1-f$ goes to zero as a positive power of $1/\gamma$, though without numerical calculations the exponent, $p \sim 1/(1+\alpha^2\epsilon)$, which lies in the range between 0 and 1, cannot be predicted precisely.

The numerical coefficient of the power of $1/\gamma$, here crudely estimated to be $(\alpha/2)^p$, would also be expected to have relatively small corrections from ignoring nonzero values of $\delta$ and $E/r_0-\alpha$ and from also ignoring various departures from the power-law relations assumed above.  However, in any case the evidence seems strong that the fraction of energy radiated away goes to zero as a positive power of $1/\gamma$ when $\gamma$ is taken to infinity, with the exponent apparently being positive but less than one.

\section{Possible Reasons for the Contrary Conclusion of Sperhake, Berti, Cardoso, and Pretorius}

Now I wish to examine some of the possible reasons for the contrary conclusion of Sperhake, Berti, Cardoso, and Pretorius \cite{Sperhake:2012me}, ``An extrapolation of our results to the limit $\gamma \rightarrow \infty$ suggests that about half of the center-of-mass energy of the system can be emitted in gravitational radiation, while the rest must be converted into rest-mass and spin energy.''

First, the numerical data SBCP used only went up to $\gamma = 2.49$, which is somewhat large compared with unity, but hardly large enough to form definite conclusions about what happens when $\gamma$ is taken to infinity. For example, Sperhake, Cardoso, Pretorius, Berti, and Gonz\'{a}lez \cite{Sperhake:2008ga} found that a good fit to their head-on collision data of the fraction, which I am hereby labeling $f_0(\gamma) = \Delta E/E$, of the gravitational wave energy radiated, $\Delta E$, out of the total energy, $E$, for impact parameter $b=0$ as a function of $\gamma \equiv E/M$ (with my $M$ and $E$, not their $M$ which is the total ADM mass that is my $E$, and not their $E$ which is my $\Delta E$) is given by the Zero Frequency Limit (ZFL) formula \cite{Smarr:1977fy}, the right hand side of Eq.\ (3) in \cite{Sperhake:2008ga}, with the energy cutoff for the ZFL evaluated as $E_\infty = 0.14\pm 0.03$, say $E_\infty = 0.14$ for the numerical values below, so that Eq.\ (3) in \cite{Sperhake:2008ga} can be written as
\ba
f_0(\gamma) &=& 0.14\left(1 + \frac{1}{2\gamma^2} 
- \frac{(4\gamma^2-1)\ln{(\gamma+\sqrt{\gamma^2-1})}}{2\gamma^3\sqrt{\gamma^2-1}}\right)\nonumber \\
&=& 0.14\left(\frac{3-v^2}{2} - (1-v^2)(3+v^2)\frac{1}{4v}\ln{\frac{1+v}{1-v}}\right)\nonumber \\
&=& 0.14\sum_{n=2}^\infty \frac{8(n-1)v^{2n}}{(2n-3)(2n-1)(2n+1)}\nonumber \\
&=& 0.14\left(\frac{8}{15}v^4 + \frac{16}{105}v^6 + \frac{8}{105}v^8 + \frac{32}{693}v^{10}
+ \frac{40}{1287}v^{12} + \frac{16}{715}v^{14} +\! \cdots\!\!\right)\!\!.
\label{f0}
\ea
For $\gamma = 2.49$, this formula gives a fraction $f_0(2.49) = 0.077$ that is only 55\% as large as the $\gamma \rightarrow \infty$ limit of 0.14, which does not seem sufficient for obtaining firm conclusions about the $\gamma \rightarrow \infty$ limit.

Second, SBCP \cite{Sperhake:2012me} fit to a formula that, as $1/\gamma \rightarrow 0$, approaches a constant with a deviation proportional to $1/\gamma$, rather than to $(1/\gamma)^p$ with the smaller exponent $p = 1/(1+\alpha^2\epsilon)$ that my analysis gives.  Therefore, it is perhaps not surprising that they did not see the slower decrease of $1-f = (E - \Delta E)/E$ to zero for $\gamma \rightarrow \infty$.

In particular, SBCP fit their numerical estimates for what I am calling $f(\gamma) = \Delta E/E$, the fraction of the total energy $E$ radiated as gravitational waves of energy $\Delta E$ at the $\gamma$-dependent critical impact parameter $b = b_c = \beta_c E$, with the following formula for what I shall label as $f_{\mathrm{SBCP}}(\gamma)$, in terms of the fraction $f_0(\gamma)$ at $b=0$:
\bb
f_{\mathrm{SBCP}}(\gamma) = \frac{f_0(\gamma)}{\mathcal{R_{\mathrm{SBCP}}}(\gamma)},\ \mathcal{R}_{\mathrm{SBCP}}(\gamma) = 0.34(1-1/\gamma).
\label{fSBCP}
\ee

Let me compare different fits to the data given for initially nonrotating black holes in Table I of \cite{Sperhake:2012me} for $f(\gamma) = \Delta E/E$, which is the product of what is labeled in Table I as $K/M = (\gamma-1)/\gamma$ and as $E_{\mathrm{rad}}/K$, with their $E_{\mathrm{rad}}$ being the same as my $\Delta E$, the total gravitational wave energy radiated.  These data for $f(\gamma)$ are (here retaining all 6 digits from multiplying the 3-digit values of $K/M$ and $E_{\mathrm{rad}}/K$ in Table I, leaving rounding to fewer digits until later)
\bb
f(1.22) = 0.160\,921,\ f(1.88) = 0.298\,116,\ f(2.49) = 0.343\,252.
\label{f-values}
\ee

More relevant for my arguments that $f(\gamma)\rightarrow 1$ as $\gamma \rightarrow \infty$ are the values for
\bb
F(\gamma) \equiv 1 - f(\gamma) = M_f/E = (\mathrm{final\ rest\ mass\ }M_f)/(\mathrm{total\ energy\ }E),
\label{F(gamma)}
\ee
where $M_f$ is the final mass of the black hole that forms for impact parameter $b$ infinitesimally below the critical impact parameter $b_c$.  Table I of \cite{Sperhake:2012me} then implies
\bb
F(1.22) = 0.839\,079,\ F(1.88) = 0.701\,884,\ F(2.49) = 0.656\,748.
\label{F-values}
\ee

The SBCP fitting function $f_{\mathrm{SBCP}}(\gamma)$ of Eqs.\ (\ref{f0}) and (\ref{fSBCP}) then gives what I call
\bb
F_{\mathrm{SBCP}}(\gamma) = 
1 - \frac{0.14/0.34}{(1 - 1/\gamma)}\left(1 + \frac{1}{2\gamma^2} 
- \frac{(4\gamma^2-1)\ln{(\gamma+\sqrt{\gamma^2-1})}}{2\gamma^3\sqrt{\gamma^2-1}}\right),
\label{FSBCP}
\ee
with values
\ba
F_{\mathrm{SBCP}}(1.22) &=& 0.854\ =\ 1.018\, F(1.22),\nonumber \\ 
F_{\mathrm{SBCP}}(1.88) &=& 0.676\ =\ 0.963\, F(1.88),\nonumber \\ 
F_{\mathrm{SBCP}}(2.49) &=& 0.620\ =\ 0.944\, F(2.49),
\label{FSBCP-values}
\ea
which have an rms error relative to the numerical values of $F(\gamma)$ of about 4.0\%.

However, since my Eq.\ (\ref{1-f}) for the asymptotic behavior of $F(\gamma)$ has two unknown parameters, $\alpha$ and $\epsilon$ (ignoring weak dependencies on the other unknown dimensionless parameters such as $\delta$ and $E/r_0-\alpha$), for a fair comparison I should replace the one-parameter SBCP fitting function $f_{\mathrm{SBCP}}(\gamma) = f_0(\gamma)/\mathcal{R}_{\mathrm{SBCP}}(\gamma)$, having its one fitted constant 0.34 in $\mathcal{R}_{\mathrm{SBCP}}(\gamma) = 0.34(1-1/\gamma)$, with a two-parameter fitting function $f_{2}(\gamma) = f_0(\gamma)/\mathcal{R}_{2}(\gamma)$ with $\mathcal{R}_{2}(\gamma) = a - b/\gamma$, having two fitting parameters, $a$ and $b$.  Then a least-squares fit to the numerical data for $f(\gamma)$ given in Eq.\ (\ref{f-values}) yields $a = 0.389$ and $b = 0.406$ after rounding to 3 digits, and hence
\ba 
F_{2}(\gamma) &=& 1 - \frac{f_0(\gamma)}{(0.389 - 0.406/\gamma)}\nonumber \\
&=&1 - \frac{0.14}{(0.389 - 0.406/\gamma)}\left(1 + \frac{1}{2\gamma^2} 
- \frac{(4\gamma^2-1)\ln{(\gamma+\sqrt{\gamma^2-1})}}{2\gamma^3\sqrt{\gamma^2-1}}\right)\!.
\label{F2}  
\ea

This fit gives
\ba
F_{2}(1.22) &=& 0.8408 = 1.0021\, F(1.22),\nonumber \\ 
F_{2}(1.88) &=& 0.7018 = 0.9999\, F(1.88),\nonumber \\ 
F_{2}(2.49) &=& 0.6577 = 1.0014\, F(2.49),
\label{F2-values}
\ea
which have an rms error relative to the numerical values of $F(\gamma)$ of only about 0.14\%, a remarkably good fit, but perhaps the extreme excellency of this fit is somewhat spurious because of the numerical uncertainties of the values of $F(\gamma)$ given in Eq.\ (\ref{F-values}) derived from the SBCP Table I of \cite{Sperhake:2012me}.

In contrast to these fits that have $1/\gamma$ appearing linearly in $\mathcal{R}_{\mathrm{SBCP}}(\gamma)$ and $\mathcal{R}_{2}(\gamma)$ (though there is a term asymptotically going as $(\ln{\gamma})/\gamma^2$ in $f_0(\gamma)$ given by Eq.\ (\ref{f0}) from the right hand side of the ZFL Eq.\ (3) of \cite{Sperhake:2008ga} that originally came from Eq.\ (2.20) of  \cite{Smarr:1977fy}), the derivation in Section 3 above of the large-$\gamma$ asymptotic form given by Eq.\ (\ref{1-f}) suggests a fit by $F_p(\gamma) = A\gamma^{-p}$ with two parameters, $p \sim 1/(1+\alpha^2\epsilon)$ and $A \sim (\alpha/2)^p$.  Then a least-squares fit of the three numerical values of $\ln{F(\gamma)}$ by $\ln{A} - p \ln{\gamma}$ gives, to three digits each, $A = 0.892$ and $p = 0.343 = (0.7)^3$, or
\bb
F_p(\gamma) = 0.892\, \gamma^{-0.343}.
\label{Fp}
\ee

This fit gives $\alpha \sim 1.44$, $\epsilon \sim 0.93$ (though I would not assign much precision to these estimated values, because of the uncertainty of the extrapolation to $\gamma \gg 1$).

Then $F_p(\gamma) = 0.892\, \gamma^{-0.343}$ gives
\ba
F_{p}(1.22) &=& 0.833 = 0.993\, F(1.22),\nonumber \\ 
F_{p}(1.88) &=& 0.718 = 1.023\, F(1.88),\nonumber \\ 
F_{p}(2.49) &=& 0.652 = 0.993\, F(2.49),
\label{Fp-values}
\ea
which has an rms relative error of about 1.46\%, a bit over 10 times the rms relative error of $F_2(\gamma)$.  However, since the numerical error of the data for $F(\gamma)$ in Eq.\ (\ref{F-values}) is likely to be a few percent, it seems that both formulas fit the three data points of Eq.\ (\ref{F-values}) quite well.  Certainly Eq.\ (\ref{Fp}) for $F_p(\gamma)$ looks simpler than Eq.\ (\ref{F2}) for $F_2(\gamma)$, and appears even simpler than Eq.\ (\ref{FSBCP}) for $F_{\mathrm{SBCP}}(\gamma)$ that does not fit quite so well, though since the complication arises mostly from the complexity of Eq.\ (\ref{f0}) for $f_0(\gamma)$, the fraction of energy radiated in a head-on collision, which is theoretically determined by the Zero Frequency Limit of  \cite{Smarr:1977fy} with only the one free parameter $E_\infty$ from the frequency cutoff, there is actually only one free parameter that is fit to the data in Eq.\ (\ref{FSBCP}), namely the fraction $0.14/0.34 = 7/17 \approx 0.412$.

The main point is that Eq.\ (\ref{1-f}) has the theoretical justification given in Section 3, as an approximate asymptotic formula for large $\gamma = E/M = (\mathrm{total\ energy})/(\mathrm{rest\ mass})$, whereas the factors of $1 - 1/\gamma$ and $0.389 - 0.406/\gamma$ in Eqs.\ (\ref{FSBCP}) and (\ref{F2}) appear to be rather {\it ad hoc}.  And even if Eq.\ (\ref{1-f}) is not quite the correct asymptotic form for the fraction of energy not radiated when $\gamma \gg 1$, the fact that during most of the evolution the black holes are expected to be much smaller than their separations very strongly suggests that they cannot absorb a large fraction of the initial energy.

A third possible reason for the disputed conclusion of Sperhake, Berti, Cardoso, and Pretorius \cite{Sperhake:2012me}, that $F = 1-f = (E - \Delta E)/E$ remains bounded below by a positive number when $\gamma$ is taken to infinity, is that the 2007 results of Pretorius and Khurana \cite{Pretorius:2007jn} suggest that $f = \Delta E/E$, the fraction of the initial total energy $E$ going into the energy $\Delta E$, is a very sensitive function of the impact parameter $b$ when it is near the critical impact parameter $b_c$.  Therefore, it might be the case that for the crucial data SBCP calculated at $\gamma = 2.49$, which was done with impact parameter $b = 2.749 E$ (in my notation; they use $M$ for the total ADM mass that I call $E$), might not have been near enough to the impact parameter $b_m = \beta_m E$ that would maximize the fraction $f = \Delta E/E$ going into gravitational waves at that value of $\gamma = E/M$ (where my $M$ is the sum of the initial two rest masses of the black holes).

An observation supporting the hypothesis that with the limited computational resources available, SBCP were not able to tune the impact parameter close enough to the value $b_m$ that would give the maximum energy radiated to correctly evaluate what that maximum is for each $\gamma$, is the fact that Figure 1 of \cite{Sperhake:2012me} shows that the example they calculated for their maximum $\gamma$, $\gamma = 2.49$, does not have the black hole separation decreasing monotonically before they reach their minimum separation and then move apart.  I would expect that at $b = b_m$, and also at $b = b_c$, the critical impact parameter that maximizes $f = \Delta E/E$ out of all critical impact parameters, the separation between the two black holes would decrease monotonically until the black holes merge.  Therefore, Figure 1 seems to be evidence that SBCP were not able to tune the impact parameter sufficiently precisely to obtain the maximum fraction $f$ of energy radiated.  Of course, this is not a criticism of their heroic effort to learn what they could at as large a value of $\gamma$ that was feasible with their calculational resources, but it does suggest that even more resources will be needed to find numerical evidence convincingly suggesting whether or not the final black hole (if the impact parameter $b$ is infinitesimally smaller than the critical impact parameter $b_c = \beta_c E$ for fixed total energy $E$ that maximizes $f = \Delta E/E$ at that $E$) can have a mass $M_f$ such that $M_f/E \rightarrow 0$ as $M/E \rightarrow 0$.

Actually, the `zoom-whirl' behavior of the two black holes SBCP calculated for $\gamma \equiv E/M = 2.49$ (giving $v \approx 0.9158$) and impact parameter $b = 2.755 E \approx 1.009 E(2.5/v)$ (incidentally fitting very closely to the estimate of Shibata, Okawa, and Yamamoto \cite{Shibata:2008rq} that $b_c \approx 2.5 E/v$) suggests to me that perhaps that value of $b$ SBCP used is closer to a critical dimensionless impact parameter that is different from the one that maximizes $f = \Delta E/E$ and which I would expect would lead to something close to an equiangular spiral in which the separation of the two black holes decreases monotonically until they merge (if $b$ is infinitesimally below the $b_c$ that maximizes $f$).  So far the data gives me only a weak hint of the possibility that there may be more than one critical impact parameter, and that the fraction $f = \Delta E/E$ of energy radiated into gravitational waves might not have a single local maximum as a function of $\gamma = E/M$.  However, it would be very interesting to explore this intriguing possibility.

I might initially {\it guess} that the number $N$ of local maxima of $f$ might grow approximately linearly with $\gamma$, and that the number of critical impact parameters might be $2N-1$ (say if there is one local maximum of $f$ for each odd-numbered critical impact parameter).  This would suggest that for fixed total energy $E \gg M$ and fixed initial total rest mass $M = E/\gamma$, although the global maximum for $f$ would occur for $b = \beta_m E \sim E$, $f(b)$ might have some oscillatory structure with a period of the order of $\delta b \sim M = E/\gamma$.  On the other hand, this would give a range for all $2N-1$ critical impact parameters of $\Delta b \sim N\delta b \sim E$, whereas I would expect the ratio of the range of the oscillatory behavior to the value of $b_m$ (with $b_m$ giving the global maximum for $f$ for the fixed values of $E$ and $M$) to decrease as $\gamma$ is increased, so perhaps the typical period of the oscillations, say $\delta b$, goes as $b_m \sim E$ multiplied by a power of $1/\gamma = M/E$ that is larger than one, so that $\delta b$ becomes far smaller than $M$ for large $\gamma$.  Or, perhaps the number $N$ of local maxima of $f$ might increase slower than linearly with $\gamma$, so that the typical period of the oscillations, $\delta b$, could remain of the order of $M$ but the range, $\Delta b \sim N\delta b$, could become much smaller than $b_m$ for large $\gamma$.  In any case, it would be very interesting to see what the behavior of $f(\gamma)$ is as a function of $\gamma = E/M$ when it becomes large, and to check whether it exhibits any oscillations, giving more than one local maximum.

Perhaps when the computational resources become available for making many more calculations at slightly different impact parameters for the same $\gamma$, and ideally also for larger $\gamma$, it might become more apparent whether or not the fraction $F$ of the total energy that does not get radiated does approach zero as $\gamma$ is increased indefinitely with $\beta \equiv b/E$ tuned to the $\gamma$-dependent value $\beta_m(\gamma)$ that maximizes the fraction $f = 1-F$ of the total energy that is radiated for each $\gamma = E/M$.

Another argument for $F(\gamma) \rightarrow 0$ as $\gamma \rightarrow \infty$ is that in the limit $1/\gamma = 0$ at fixed total energy $E$, the rest mass of each black hole, $M/2 = E/(2\gamma)$, goes to zero, so initially one has two incoming Aichelburg-Sexl metrics \cite{Aichelburg:1970dh}, corresponding to the gravitational fields of two classical massless point particles.  In this case there are, at least initially, no black holes to absorb any of the energy, so when SBCP argue \cite{Sperhake:2012me} that ``absorption sets an
upper bound on the maximum energy that can be radiated,'' it would seem that no absorption would occur in the $M=0$ limit, since there are no black holes to absorb the energy (until the two massless particles form a single black hole).

One weakness of this last argument is that it might occur that in the Aichelburg-Sexl limit, as massless particle A collides with the gravitational shock wave of massless particle B, the gravitational shock wave field of A might focus the gravitational shock wave field of B to form a black hole separate from a black hole that might form from the field of B focusing the shock wave of A.  Thus {\it perhaps} two black holes {\it might} form that conceivably each could have its rest mass be a fraction of the total energy that is bounded below by a positive number.  In this case that positive number would give a lower bound on the ratio $M_f/E$ for the final black hole that forms with impact parameter $b$ slightly below the critical impact parameter $b_c$.

This conceptual possibility raises the question of whether these possible two distinct holes each engulf the corresponding massless particle, which might limit the number of black holes that form to that pair, or whether the two holes that might form from the focusing of the gravitational shock wave of one massless particle by the shock wave of the other might form behind the massless particles rather than engulfing them.  This latter possibility raises the question of whether the gravitational fields of the two persisting massless particles might focus to produce even more pairs of black holes.  Although I am rather sceptical that the gravitational fields of the two massless particles can form even one pair of black holes that persist before merging to form one final black hole (if $b < b_c$) or eventually flying apart (if $b > b_c$), if two black holes can form without engulfing the massless particles, it would seem conceivable that an arbitrarily large number of pairs of black holes could form from suitable tuning of $\beta = b/E$, and even that an arbitrarily large number could survive the encounter rather than merging if $b < b_c$.

However, the more plausible possibility for $1/\gamma = M/E = 0$ seems to be that the two massless particles that are the source of the original pair of Aichelburg-Sexl metrics cannot form separate black holes but only a single black hole if they merge.  In this case, there are never any separate black holes to absorb any of the radiation, and it seems most plausible that by fine tuning the ratio $\beta = b/E$ of the impact parameter $b$ to the total energy $E$ to the critical value $\beta_0$, all of the energy would be radiated away before the two particles merge and disappear at a naked singularity corresponding to a final single black hole in the limit that its mass goes to zero.

\section{Conclusions}

In this paper I have given arguments that the 2007 conjecture of Pretorius and Khurana \cite{Pretorius:2007jn} is essentially correct, that all but an arbitrarily small fraction of the total energy $E$ of ultrarelativistic objects of original total rest mass $M$ can be converted to gravitational waves if $M/E$ can be taken small enough.  (I do not argue that {\it all} of the kinetic energy can be converted, since the final total rest mass $M_f$ could be larger than $M$, but only that as $M/E$ is taken to zero, $M_f/E$ also approaches zero with the proper fine tuning of the impact parameter.)  The argument for two original black holes with $M/E \ll 1$ is that during most of the gravitational wave emission, the black holes are so much smaller than the region where most of the gravitational wave energy is located that they can absorb only a tiny fraction of it, with this fraction going to zero as $M/E$ is taken to zero.

Contrary arguments in 2012 by Sperhake, Berti, Cardoso, and Pretorius do not seem sufficient to refute this argument, since (1) they could only calculate the emission near the critical impact parameter for $\gamma \leq 2.49$, (2) they used a rather {\it ad hoc} fitting function that does not agree with the theoretical expectations put forward in this paper, and (3) they do not seem to have been able to tune the impact parameter close enough to the critical value to get a convincingly good estimate of the maximum fraction of energy that could be emitted for their finite values of $E/M$.

\section*{Acknowledgments}

I am grateful for email discussions with Emanuele Berti, Vitor Cardoso, Frans Pretorius, Ulrich Sperhake, Gabriele Veneziano, and Huan Yang.  This work was supported in part by the Natural Sciences and Engineering Research Council of Canada.


\end{document}